\documentclass[twocolumn,showpacs,aps,prl]{revtex4}
\usepackage{graphicx}
\usepackage{dcolumn}
\usepackage{amsmath}

\begin{document}

\title{Eccentricity fluctuations and its possible effect on
  elliptic flow measurements}

\author{
Mike~Miller$^{(a)}$ and 
Raimond~Snellings$^{(b)}$
} 

\affiliation{ 
(a) Yale University, New Haven, Connecticut 06520, USA \\
(b) NIKHEF, Kruislaan 409, 1098 SJ Amsterdam, The Netherlands
}

\date{\today}

\begin{abstract}
The elliptic flow measured at RHIC has been interpreted as a
signature for strong partonic interactions early in the collision and
as an indication of a well developed quark-gluon plasma phase. 
The measured values of elliptic flow, 
using methods based on multi-particle
correlations, are affected by fluctuations in the magnitude of the
elliptic flow.
In this Letter, using a Monte Carlo Glauber calculation, we 
estimate what the possible effect of spatial 
eccentricity fluctuations is on the determination of elliptic flow.
\end{abstract}


\maketitle

\begin{figure}[thb]
 \begin{center}
   \includegraphics[width=0.3\textwidth]
   {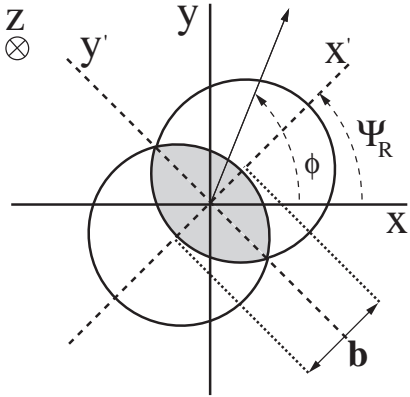}
   \caption{
     Schematic view of a nucleus-nucleus collision in the transverse
     plane. 
    }
    \label{geometry}
 \end{center}
\end{figure}

In non-central heavy-ion collisions, the initial spatial anisotropy
due to the geometry of the overlap region and the pressure developed
early in the collision generate an observable azimuthal momentum-space
anisotropy. 
The particle yields produced in heavy-ion
collisions can be characterized by~\cite{VoloshinFourier}:
\begin{equation}
  \frac{{\rm d}^3N}{{\rm d}p^2_t{\rm d}\phi{\rm d}y} = 
  \frac{{\rm d}^2N}{2\pi{\rm d}p^2_t{\rm d}y} [1 + 2\sum_n{v_n
    {\rm cos}(n(\phi - \Psi_R))}],
\end{equation}
where $p_t$ is the transverse momentum of the particle, $\phi$ is
its azimuthal angle, $y$ is the rapidity and $\Psi_R$ the reaction
plane angle, see fig~\ref{geometry}. 
The second coefficient, $v_2$, of this  
Fourier series is called {\it elliptic flow}.

Elliptic flow as a signature of hydrodynamic behavior of nuclear
matter produced in high energy nuclear collisions has been proposed
by Ollitrault in 1992~\cite{Ollitrault92}. 
After that it has attracted increasing attention
from both experimentalists and theorists~\cite{OverviewFlowTheory} 
and has been measured at AGS~\cite{E877,E895}, SPS~\cite{NA49,CERES} 
and RHIC~\cite{STAR,PHOBOS,PHENIX} energies. 
It is thought that elliptic flow reflects the
amount of interactions between the constituents at an early time in the
evolution of the produced system~\cite{SorgeEarly}. 
Therefore it is sensitive to the
equation of state of the produced system when this system might be in
the quark-gluon plasma phase.   

Since the reaction plane is not known experimentally, the elliptic
flow is calculated using azimuthal angular correlations between the
observed particles~\cite{PoskanzerVoloshinMethods}. 
In the case of two particle correlations the measurement is
proportional to $v_2^2$. 
The reported elliptic flow values are therefore obtained as
$\sqrt{\langle v_2^2 \rangle}$ after averaging over events.  

Because elliptic flow is a collective effect, it is a correlation of
all the particles with the reaction plane. This can be exploited
experimentally by
using multiple particle correlations to calculate $v_2$. 
To calculate these correlations a convenient
mathematical approach is to use cumulants. This method, proposed
in~\cite{OllitraultCumulants}, has the additional advantage that it
allows to subtract the so called non-flow effects from $v_2$. 
Non-flow effects are
correlations between the particles not related to the reaction plane.
Such effects include, but are not limited to, resonance decays,
(mini)jet fragmentation and Bose-Einstein correlations. 
The cumulant method uses multi-particle correlations which introduce 
higher powers of $v_2$. The corresponding equations for calculating
$v_2$ in the cumulant method for two, four and six particle
azimuthal correlations are given by: 
\begin{eqnarray}
    (v_2\{2\})^2 &=& \langle v_2^2 \rangle \nonumber \\
    (v_2\{4\})^4 &=& 2 \langle v_2^2\rangle^2 - \langle
    v_2^4\rangle \nonumber \\
    (v_2\{6\})^6 &=& \frac{1}{4}\left(\langle v_2^6\rangle -9\langle v_2^4\rangle
    \langle v_2^2\rangle + 12 \langle v_2^2\rangle^3\right),
    \label{fcummu}
\end{eqnarray}
where $v_2\{2\}$ is calculated using two-particle azimuthal
correlations, $v_2\{4\}$ using a mix of two and four-particle
azimuthal correlations and $v_2\{6\}$ uses two, four and six-particle
azimuthal correlations. 

However, due to event-by-event fluctuations in the
elliptic flow for instance, the event averaged $\langle v_2^n \rangle
\neq \langle v_2 \rangle^n$ for $n \ge 2$~\cite{STAR_flowPRC}. 
Therefore, comparing the experimental values of
$v_2$ with model calculations which do not correctly include these
fluctuations might not be {\it a priory} justified. 
Furthermore, there are effects due to the 
finite width of the centrality bins
where also $\langle v_2^n \rangle \neq \langle v_2 \rangle^n$. The STAR
collaboration~\cite{STAR_flowPRC} found a negligible bias on the
extracted $v_2$ due to these binning effects, except in the most
central bin.  

\begin{figure}[t]
 \begin{center}
   \includegraphics[width=0.5\textwidth]
   {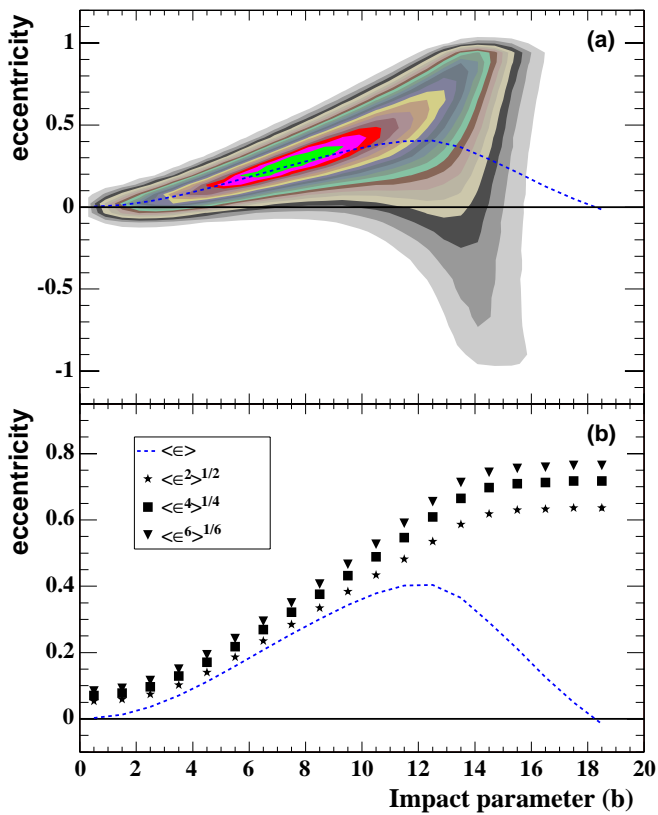}
   \caption{
      a) Contour plot of the calculated eccentricity in a Monte Carlo
      Glauber model versus the impact parameter, {$b$}, in Au+Au
      collisions.  The mean value of the eccentricity is indicated by
      the dashed curve.
      b) The mean eccentricity $\langle \epsilon \rangle$ (dashed
      curve) and the corresponding 
      $\langle \epsilon^{n} \rangle^{1/n}$ 
      for $n = 2$, 4 and 6 (points) versus impact parameter $b$. 
    }
    \label{fig1}
 \end{center}
\end{figure}

To estimate the possible effect of the fluctuations on the
measured elliptic flow values we calculate the fluctuations in the
initial {\it spatial} anisotropy of the created system using a Monte
Carlo Glauber model (MCG).
This anisotropy, which generates the elliptic
flow, is given by~\cite{SorgeEpsilon, Heiselberg}:
\begin{equation}
  \epsilon \equiv \frac{\sum y'^{2}_{i}  - \sum x'^{2}_{i}}
  {\sum y'^{2}_{i} + \sum x'^{2}_{i}},
  \label{epsilon}
\end{equation}
where $x'_i$ and $y'_i$ are the coordinates of the constituents in the
plane perpendicular to the beam and $x'$ is in the reaction plane (see
Fig.~\ref{geometry}).  
Due to the relation between the initial spatial anisotropy and the
elliptic flow, $v_2$ $\propto \epsilon$~\cite{Heiselberg, 
  SorgeEpsilon, PoskanzerVoloshinEpsilon}, fluctuations in
$\epsilon$ will lead to fluctuations in $v_2$. 
It should be noted that this is only one specific example of
fluctuations which could contribute to the measured
value of $v_2$. 

The MCG approach allows
for an event-by-event calculation of $\epsilon$ and therefore the
determination of the corresponding 
higher order moments of the event averaged distribution of $\epsilon$. 
For details on the MCG model and the parameters used see
Ref.~\cite{STAR_pionGlauber}. 

In Fig.~\ref{fig1}a, the eccentricity calculated using the MCG is
plotted versus the impact parameter ($b$) of an Au+Au collision.
This figure shows that the fluctuations in $\epsilon$ for the
most peripheral collisions are large. For the most central collisions
$\epsilon$ can be both positive and negative. This leads to an obvious
bias when calculating $\epsilon$ from $\langle \epsilon^2
\rangle^{1/2}$.
Figure~\ref{fig1}b shows the calculated $\langle \epsilon \rangle$ and
the corresponding $\langle \epsilon^{n} \rangle^{1/n}$. 
It is clear that there is a bias over the whole
centrality range, however this bias is the largest for the most
central and most peripheral collisions.

\begin{figure}[tb]
 \begin{center}
   \includegraphics[width=0.5\textwidth]
   {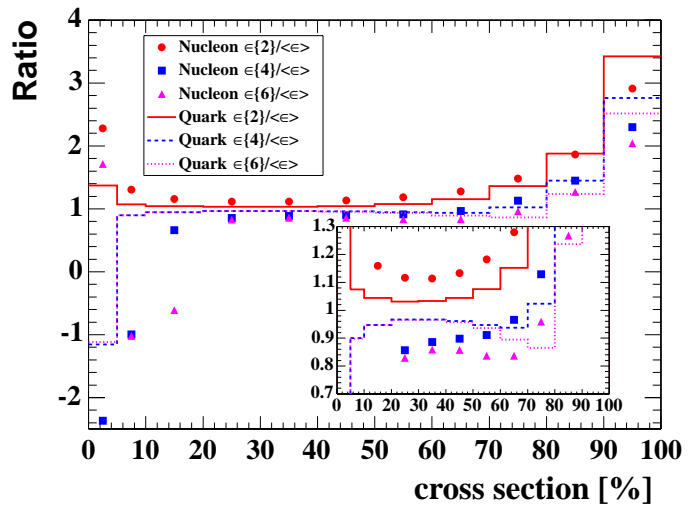}
   \caption{
     Eccentricity cumulants calculated for Au+Au collisions in a
     Monte Carlo Glauber model compared to $\langle \epsilon
     \rangle$ as a function of centrality (from left to right is from
     central to peripheral collisions, respectively). 
     The symbols are the Monte Carlo Glauber results using nucleons, the
     lines are for constituent quarks (see text). 
    }
    \label{fig2}
 \end{center}
\end{figure}

Using $v_2 \propto \epsilon$, we replace in Eq.~\ref{fcummu} $v_2$ 
by $\epsilon$, which allows us to calculate $\epsilon\{2\}$,
$\epsilon\{4\}$ and $\epsilon\{6\}$.
If the right-hand side of Eq.~\ref{fcummu} becomes negative, we take
the $n^{\rm th}$ root of the absolute value and multiply this by $-1$.
Figure~\ref{fig2} shows the ratio of $\epsilon\{m\}$ and $\langle
\epsilon \rangle$, where $m =2,4,6$, versus the collision centrality
in terms of cross-section. 
In the standard MCG approach for nucleons with a cross-section of 42 mb
the event by event fluctuations lead to a ratio $\epsilon\{2\}/\langle
\epsilon \rangle$ which is always larger than 1. This shows that the
experimental determination of $v_2$, using two-particle azimuthal
correlations in the case of event-by-event fluctuations in $\epsilon$,
leads to an overestimation of the true elliptic flow. This is
particularly true for the most central and most peripheral events.
The inset in Fig.~\ref{fig2} shows the same calculated values on a expanded
scale. Even for the centrality region of 20-60\% the
two-particle correlation method overestimates the true elliptic flow by
about 10\%. In the same figure the results for the higher order
cumulant ratios, $\epsilon\{4\}/\langle \epsilon \rangle$ and
$\epsilon\{6\}/\langle \epsilon \rangle$, are shown. The higher order
cumulants in the case of event by event fluctuations underestimate
the true elliptic flow in the centrality range of 0-80\%. Above 80\% of
the cross section also the higher order cumulants overestimate the
elliptic flow. In the mid-central region (20-60\%)
the higher order cumulants underestimate the true elliptic flow by
about 10\%. If the fluctuations are small it can be shown analitically
that the true value of $v_2$ is given by $v_2\{2\} + v_2\{4\} /
2$~\cite{Ollipriv}. 

The most central 10\% of the cross section show an interesting
behavior for the different cumulants. The two particle correlations
always lead to a real value of $\epsilon\{2\}$, however the higher
order cumulants can become complex because of the combination of
two, four and higher multi-particle contributions. This is the case
for $\epsilon\{4\}$ and $\epsilon\{6\}$ in the centrality bin 5-10\%.
However, in the most central bin, 0-5\%, $\epsilon\{6\}$ becomes
real again. Complex values of $v_2$ are
usually not reported by experiments. 

To investigate the sensitivity to the magnitude of the fluctuations in
the eccentricity we followed the work of Eremin and
Voloshin~\cite{VoloshinQuarkPart} and calculate $\epsilon\{m\}$ using
valence quarks as constituents in the MCG. 
The cross section used for the quarks was set to 6 mb and the number of
constituent quarks in a gold nucleus to 591. All the other parameters
were kept the same as in the nucleon MCG.  
Due to the larger number of interacting constituents the fluctuations
in central collisions are reduced.
Indeed, the $\epsilon$ fluctuations for the 0-80\% of the
cross-section are reduced as shown by the lines in Fig.~\ref{fig2}.

\begin{figure}[tb]
 \begin{center}
   \includegraphics[width=0.5\textwidth]
   {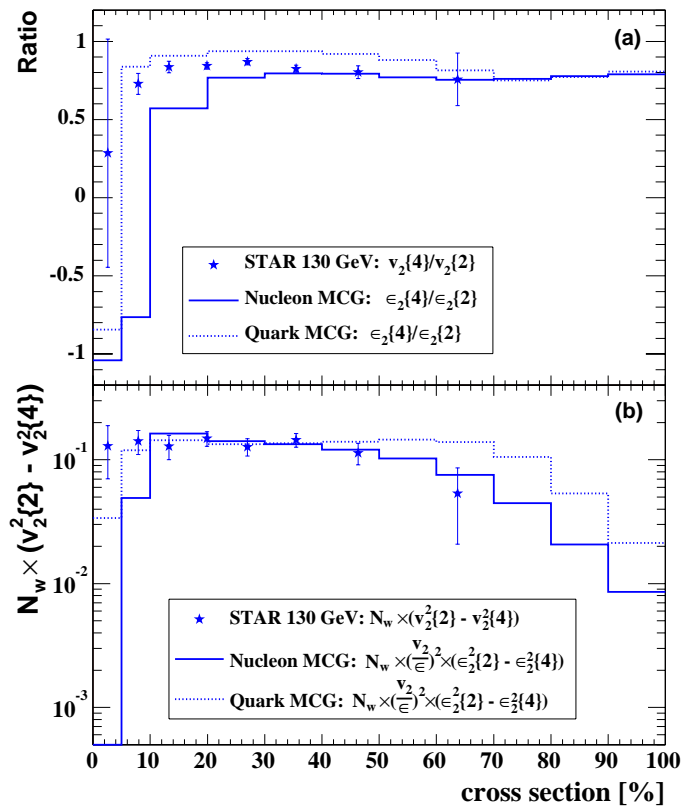}
   \caption{
      a) Comparison of the calculated $\epsilon\{4\}/\epsilon\{2\}$ for
      both the nucleon (solid line) and quark (dashed line) MCG to the
      STAR measurements~\cite{STAR_flowPRC}.
      b) Non-flow or fluctuation contribution to the
      azimuthal correlations (see text). 
    }
    \label{fig3}
 \end{center}
\end{figure}

The ratios shown in Fig.~\ref{fig2} cannot be directly compared to the
elliptic flow measurements since $\langle v_2 \rangle$ is not
an experimental observable.
Instead, we calculate the ratio $\epsilon\{m\}/\epsilon\{n\}$. 
Because of the assumed proportionality between $\epsilon$ and $v_2$, 
which in principle could depend on centrality~\cite{STAR_flowPRC}, 
this ratio can be compared with the measured $v_2\{m\}/v_2\{n\}$.

Figure~\ref{fig3}a shows the ratio
$\epsilon\{4\}/\epsilon\{2\}$ for the quark and nucleon MCG
calculations. In both cases this ratio is smaller than unity
over the whole centrality range, with the largest suppression in the
case of the nucleon MCG. The elliptic flow cumulant measurements from
the STAR collaboration~\cite{STAR_flowPRC} are also shown in
Fig.~\ref{fig3}. The experimental results are in between the calculated
values, and are closer to the nucleon (quark) MCG results for
peripheral (central) collisions.

If the observed difference between $v_2\{2\}$ and $v_2\{4\}$ would be
due to non-flow effects, 
it is proposed in Ref.~\cite{OllitraultCumulants} that the squared
difference is given by
\begin{equation}
v_2^2\{2\} - v_2^2\{4\} = \frac{g_2}{N}
\end{equation}
where $N$ is the multiplicity and $g_2/N$ is the non-flow
contribution. 
NA49~\cite{NA49flow} has indeed observed such a constant behavior.

This squared difference, as measured by the STAR
collaboration~\cite{STAR_flowPRC}, is shown in Fig.~\ref{fig3}b for
the cumulant measurements as a function of centrality.
Instead of scaling with the multiplicity as done by NA49 we scaled
the STAR results using the corresponding number of wounded
nucleons ($N_W$), which is proportional to the multiplicity.
These measurements are compared with the values obtained from the quark
and nucleon MCG (histograms in Fig.~\ref{fig3}b). 
However, to make this comparison the proportionality between
$v_2$ and $\epsilon$ was assumed to be 0.16, independent of centrality.
It is seen from the figure that the centrality dependence of the
data is better described by the nucleon MCG than by a
constant value as expected for non-flow contributions. 

In summary, experimental measurements of elliptic flow ($v_2$) might
be affected by fluctuations. In this study we investigate the effect
on the determination of $v_2$ of
fluctuations in the spatial anisotropy of the overlap
region ($\epsilon$) in heavy-ion collisions using a Monte Carlo
Glauber simulation. For this study we assume a proportional relationship
between $\epsilon$ and $v_2$. 

When elliptic flow is measured from two-particle correlations
($v_2\{2\}$) we find that fluctuations in $\epsilon$ lead to an
overestimation of its value. On the
other hand a cumulant analysis of
order four ($v_2\{4\}$) and six ($v_2\{6\}$) leads to an
underestimation of the true value. 

Experimental results on the ratios $v_2\{m\}/v_2\{n\}$ can be compared
directly to the model and are shown to be in approximate agreement.  
Traditionally, a ratio different from unity is interpreted as being
due to non-flow effects. However, this study shows that this difference 
can also be explained by event-by-event fluctuations in the elliptic
flow. In reality it is likely that both non-flow and fluctuations
affect the measured elliptic flow. It remains to be determined
which effect dominates. 

Experimentally, this can be addressed by also reporting the
ratios $v_2\{m\}/v_2\{n\}$ obtained with the cumulant analysis as a
function of centrality when they become complex because this is
a unique feature of fluctuations.
Event-by-event fluctuations in elliptic flow have been suggested
as an interesting physics observable~\cite{otherFluct}, however to
address this these  more ``trivial'' causes of fluctuations have to be
understood first. 
\\
\noindent {\bf Acknowledgments:}

We are grateful to Sergei~Voloshin, Aihong~Tang, Art~Poskanzer, 
Jean-Yves~Ollitrault, Peter~Kolb, Pasi~Huovinen and Michiel~Botje for
useful discussions and suggestions.


\begin{thebibliography}{99}

\bibitem{VoloshinFourier}
S.~Voloshin and Y.~Zhang, Z.~Phys.~C {\bf 70}, 665 (1996).

\bibitem{SorgeEpsilon}
H.~Sorge, Phys.~Rev.~Lett. {\bf 82}, 2048 (1999).

\bibitem{Heiselberg} 
H.~Heiselberg and A.-M.~Levy, Phys. Rev.~C{\bf 59}, 2716 (1999).

\bibitem{Ollitrault92} 
J.-Y.~Ollitrault, Phys. Rev.~D {\bf 46}, 229 (1992).

\bibitem{OverviewFlowTheory}
Peter~F.~Kolb and Ulrich~Heinz, review for 'Quark Gluon Plasma 3', 
nucl-th/0305084.
Pasi~Huovinen, review for 'Quark Gluon Plasma 3', nucl-th/0305064.
D.~Teany, J.~Lauret and E.~V.~Shuryak, nucl-th/0110037; 
Phys.~Rev.~Lett {\bf 86}, 4783 (2001). 
 
\bibitem{E877}
E877 Collaboration, J.~Barrette {\it et al.}, Phys. Rev. Lett. {\bf 73}, 2532 (1994).

\bibitem{E895} 
E895 Collaboration, H.~Liu {\it et al.}, Nucl. Phys. {\bf A638}, 451c
(1998).

\bibitem{NA49}
NA49 Collaboration, H.~Appelsh\"auser {\it et al.},
Phys. Rev. Lett. {\bf 80}, 4136 (1998).

\bibitem{NA49flow}
NA49 Collaboration, C.~Alt {\it et al.}, 
Phys.~Rev.~C {\bf 68}, (2003) 034903.

\bibitem{CERES}
CERES/NA45 Collaboration, G.~Agakichiev {\it et al.}, nucl-ex/0303014.

\bibitem{STAR}
STAR Collaboration, K.H.~Ackermann {\it et al.}, Phys.~Rev.~Lett. {\bf
  86}, 402 (2001).
STAR Collaboration, C.~Adler {\it et al.}, Phys.~Rev.~Lett. {\bf 87},
182301 (2001).
STAR Collaboration, C.~Adler {\it et al.}, Phys.~Rev.~Lett. {\bf 89},
132301 (2002).
STAR Collaboration, C.~Adler {\it et al.}, Phys.~Rev.~Lett. {\bf 90},
032301 (2003).
STAR Collaboration, J.~Adams {\it et al.}, nucl-ex/0306007.

\bibitem{PHOBOS}
PHOBOS Collaboration, B.B.~Back {\it et al.}, Phys.~Rev.~Lett. {\bf
  89}, 222301 (2002).

\bibitem{PHENIX}
PHENIX Collaboration, K.~Adcox {\it et al.}, Phys.~Rev.~Lett. {\bf
  89}, 212301 (2002).
PHENIX Collaboration, S.S.~Adler {\it et al.}, nucl-ex/0306021.

\bibitem{SorgeEarly}
H.~Sorge, Phys.~Rev.~Lett. {\bf 78}, 2309 (1997).

\bibitem{PoskanzerVoloshinMethods}
A.M.~Poskanzer and S.A.~Voloshin, Phys.~Rev.~C {\bf 58}, 1671 (1998).

\bibitem{OllitraultCumulants} 
Nicolas~Borghini, Phuong~Mai~Dinh, Jean-Yves~Ollitrault, Phys.~Rev.~C
{\bf 63}, 054906 (2001).
Nicolas~Borghini, Phuong~Mai~Dinh, Jean-Yves~Ollitrault, Phys.~Rev.~C
{\bf 64}, 054901 (2001).
N.~Borghini, P.M~Dinh, J.Y.~Ollitrault, Phys.~Rev.~C
{\bf 66}, 014905 (2002).
R.S.~Bhalerao, N.~Borghini, J.-Y.~Ollitrault, nucl-th/0307018.

\bibitem{STAR_flowPRC}
STAR Collaboration, C.~Adler {\it et al.}, Phys.~Rev.~C {\bf 66},
061901 (2002).

\bibitem{PoskanzerVoloshinEpsilon}
S.A.~Voloshin and A.M.~Poskanzer, Phys.~Lett. {\bf B474}, 27 (2000).

\bibitem{STAR_pionGlauber}
STAR Collaboration, J.~Adams {\it et al.}, nucl-ex/0311017

\bibitem{Ollipriv}
J.-Y.~Ollitrault, private communication (2003).

\bibitem{VoloshinQuarkPart}
S.~Eremin and S.~Voloshin, Phys.~Rev.~C {\bf 67}, 064905 (2003).

\bibitem{otherFluct}
Stanislaw~Mrowczynski and Edward~V.~Shuryak, Acta Phys. Polon. B34,
4241 (2003).

\end{thebibliography}
\end{document}